\documentclass[reprint,aps, prl,superscriptaddress]{revtex4-1}

\usepackage{amsmath, amssymb, times, mathrsfs, hyperref, array, bbm}
\usepackage{graphicx}
\usepackage[usenames,dvipsnames]{xcolor}
\usepackage{braket}

\hypersetup{colorlinks=false}

\bibpunct{[}{]}{,}{n}{}{}
\usepackage{ulem}

\renewcommand{\vec}[1]{{\boldsymbol #1}}

\begin{document}
\title{Superconducting diode effect due to magnetochiral anisotropy\\in topological insulator and Rashba nanowires}

\author{Henry F. Legg}
\affiliation{Department of Physics, University of Basel, Klingelbergstrasse 82, CH-4056 Basel, Switzerland}

\author{Daniel Loss}
\affiliation{Department of Physics, University of Basel, Klingelbergstrasse 82, CH-4056 Basel, Switzerland}

\author{Jelena Klinovaja}
\affiliation{Department of Physics, University of Basel, Klingelbergstrasse 82, CH-4056 Basel, Switzerland}

\begin{abstract}
The critical current of a superconductor can depend on the direction of current flow due to magnetochiral anisotropy when both inversion and time-reversal symmetry are broken, an effect known as the superconducting (SC) diode effect. Here, we consider one-dimensional (1D) systems in which superconductivity is induced via the proximity effect. In both topological insulator and Rashba nanowires, the SC diode effect due to a magnetic field applied along the spin-polarization axis 
and perpendicular to 
the nanowire provides a measure of inversion symmetry breaking in the presence of a superconductor. Furthermore, a strong dependence of the SC diode effect on an additional component of magnetic field applied parallel to the nanowire as well as on the position of the chemical potential can be used to detect that a device is in the region of parameter space where the phase transition to topological superconductivity is expected to arise.
\end{abstract}

\maketitle

{\it Introduction.} Rectification is the phenomenon by which the resistance, $R$, due to a current, $I$, depends on the direction of current flow, in other words $R(+I)\neq R(-I)$. When both inversion and time reversal symmetry are broken in a material or device, an effect known as magnetochiral anisotropy (MCA) can occur~\cite{Rikken2001, Rikken2005, Pop2014, Ideue2017, Yokouchi2017, Tokura2018, He2018,Wang2022,Legg2022MCA}. In the normal state of diffusive systems, for example, MCA of the energy spectrum can result in a correction to Ohm's law at second order in current, such that $V=I R_0(1+\gamma BI)$, with $V$ the applied voltage, 
$R_0$ the reciprocal resistance, $B$ the magnetic field strength, and $\gamma$ the MCA rectification coefficient \cite{Ideue2017}. Recently, a giant MCA rectification with very large $\gamma$ was observed in the normal state of topological insulator (TI) nanowires  \cite{Legg2022MCA}.

\begin{figure}[t]
	\centering
  \includegraphics[width=1\columnwidth]{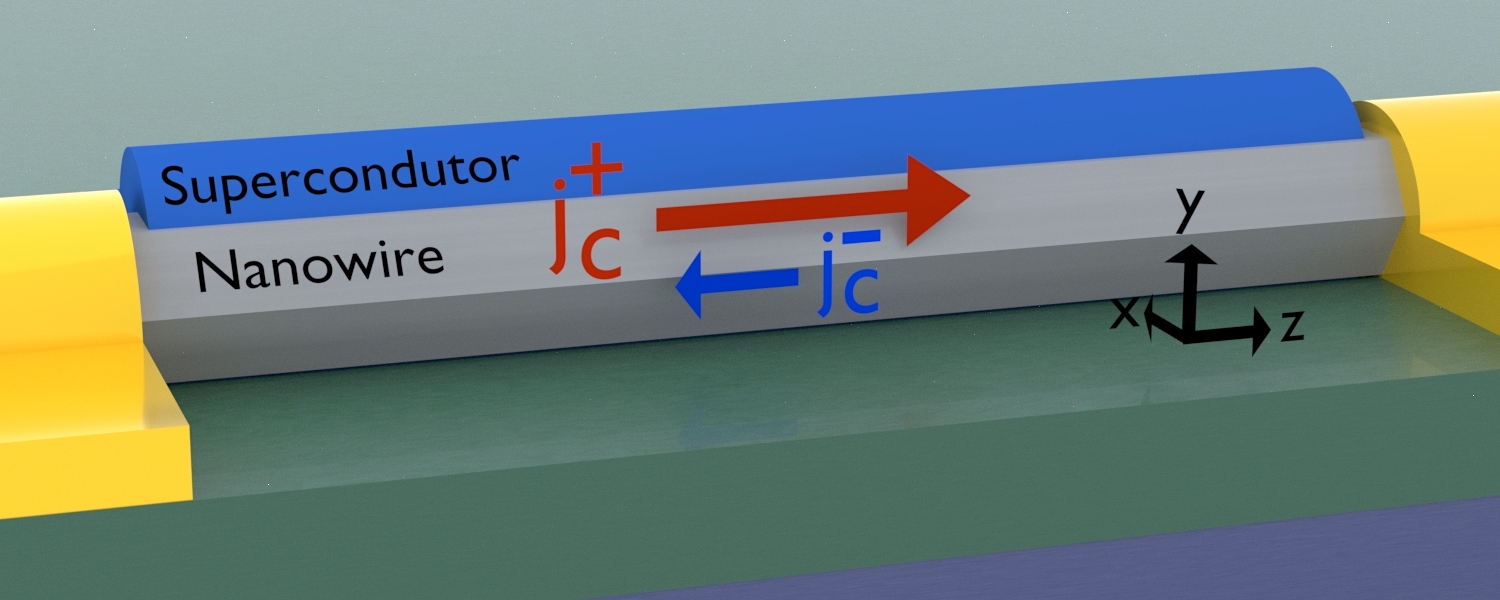}
	\caption{{\bf SC diode effect due to magnetochiral anisotropy in nanowire devices:} When the subbands of a nanowire possess a finite spin polarization due to broken inversion symmetry, a magnetic field applied along the spin-polarization direction results in a relative Zeeman shift of the subbands. The magnetochiral anisotropy (MCA) of the energy spectrum can lead to MCA rectification in the diffusive normal state~\cite{Legg2022MCA}. On the other hand, if a nanowire is brought into proximity with a superconductor, the MCA of the energy spectrum results in a critical supercurrent  in the proximitized nanowire that is different depending on whether current flows to the left or right of the device, $j_c^+\neq |j_c^-|$, the SC diode effect. The dependence of this diode effect on an additional magnetic field component parallel to the nanowire can be used to detect that the nanowire is in parameter regime where topological superconductivity is expected.}
	\label{schematic}
\end{figure}

It has been shown that normal state rectification due to MCA 
can have a counterpart in the superconducting (SC) phase \cite{Qin2017,Ando2020,Baumgartner2021}. Namely, when a current flows in a superconductor where both time reversal and inversion symmetry are broken, the critical current density at which breakdown of superconductivity occurs can be non-reciprocal too, such that $j^+_c\neq |j^-_c|$~\cite{Ando2020,Baumgartner2021,Lyu2021,Diez2021,Wu2022,He2022,Daido2022,Noah2022,Souto2022,Hou2022}, where $\pm$ indicates the direction of current flow and relative sign of  critical current (see Fig.~\ref{schematic}). This is known as the {\it SC diode} effect since, e.g. a current density $|j^-_c|<|j|<j^+_c$ experiences zero-resistance in only one direction and is in the normal state for flow in the opposite direction \cite{Ando2020,Baumgartner2021}. It has  been argued that the SC diode effect is robust against the influence of disorder~\cite{Ilic2022}.

Superconductors with broken inversion symmetry also play a key role in the search for topological superconductivity and associated Majorana bound states (MBSs) \cite{Alicea2012,Oreg2010,Lutchyn2010,Lutchyn2018,Prada2020,Legg2021,Laubscher2021}. The main experimental focus to achieve topological superconductivity has been on engineering hybrid platforms in which the superconducting pairing potential is induced in a material with the desired properties by bringing it into proximity with a trivial superconductor such as Al or Nb \cite{Lutchyn2018}.  Extensive efforts have been made to achieve topological superconductivity in one-dimensional (1D) superconductor-semiconductor nanowire devices made from materials where broken inversion symmetry is manifested as strong Rashba spin-orbit interaction (SOI), for instance in InAs and InSb~\cite{Oreg2010,Lutchyn2010,Lutchyn2018,Prada2020,Laubscher2021}. 
Another platform predicted to realize MBSs that has seen much progress recently are thin TI nanowires with quantum confined surface states \cite{Peng2010,Cook2011,hong2012,hamdou2013,cho2015,jauregui2015,jauregui2016,Ziegler2018,rosenbach2020,rosenbach2021,Munning2021,Breunig2021,Fischer2022}. In TI nanowires inversion symmetry can be broken by a non-uniform potential -- e.g. due to gating or contact with a superconductor -- and enables MBSs to appear in a large region of phase space \cite{Legg2022MCA,Legg2021,Legg2022metal}.

Despite extensive efforts to achieve MBSs in Rashba nanowires, there has been no conclusive observation so far~\cite{Prada2020} and key ingredients, such as the size of the SOI energy ($E_{\rm so}$) in the presence of the superconducting shell, the position of the chemical potential, or the values of the $g$ factor, remain largely undetermined. In fact, it has been shown that coupling a Rashba nanowire to a superconductor can renormalize its properties away from the desirable parameter range such that, for instance, $E_{\rm so}$ is reduced~\cite{reeg2017a,reeg2017,reeg2018,antipov2018,woods2018,mikkelsen2018}. In contrast to Rashba nanowires, the metallization effect due to bringing a TI nanowire into proximity with a superconductor can either enhance or reduce the size of subband-splitting that is key to achieving MBSs \cite{Legg2022metal}. However, although the giant MCA rectification recently observed in TI nanowires is evidence of a large subband-splitting in the normal state \cite{Legg2022MCA}, strongly broken inversion symmetry and the presence of a topological phase transition in superconductor-TI nanowire devices also remain undetermined experimentally.

Here we show that MCA of the energy spectrum of a TI or Rashba nanowire results in an SC diode effect when an SC pairing potential is induced by bringing the nanowire into proximity with a superconductor. The magnitude of the SC diode effect provides a measure of inversion symmetry breaking in the nanowire when the superconductor is present. Furthermore, a strong dependence of the SC diode effect on an additional magnetic field component parallel to the nanowire axis can be used as an indicator that the system parameters are close to ones at which topological superconductivity and associated MBSs are expected.  

{\it Theory of SC diode effects in nanowires:} We consider quasi-1D nanowire systems with a single occupied band, for which a normal state Hamiltonian is given by
\begin{equation}
h_k=\xi_k+\sigma_x \alpha_k+\Delta_x \sigma_x + \Delta_z \sigma_z. \label{eq:genham}
\end{equation}
Here, $k$ is momentum along the nanowire and the Pauli matrices $\sigma_i$ act in spin space, $\xi_k$ corresponds to the dispersion relation that respects inversion symmetry ($\xi_k =\xi_{-k}$), and  $\alpha_k$ corresponds to the inversion symmetry breaking term with $\alpha_k = - \alpha_{-k}$. The last terms correspond to the effective Zeeman splitting with energy $\Delta_x=g_x B_x$ and $\Delta_z=g_z B_z$ due to a magnetic field ${\vec B}=(B_x,0,B_z)$ with $g$ factors $g_x$ and $g_z$. For simplicity, throughout this manuscript, the nanowire axis is along the $z$-direction, while the direction of spin polarization due to inversion symmetry breaking is along the $x$-direction. 

\begin{figure}[t]
	\centering
  \includegraphics[width=1\columnwidth]{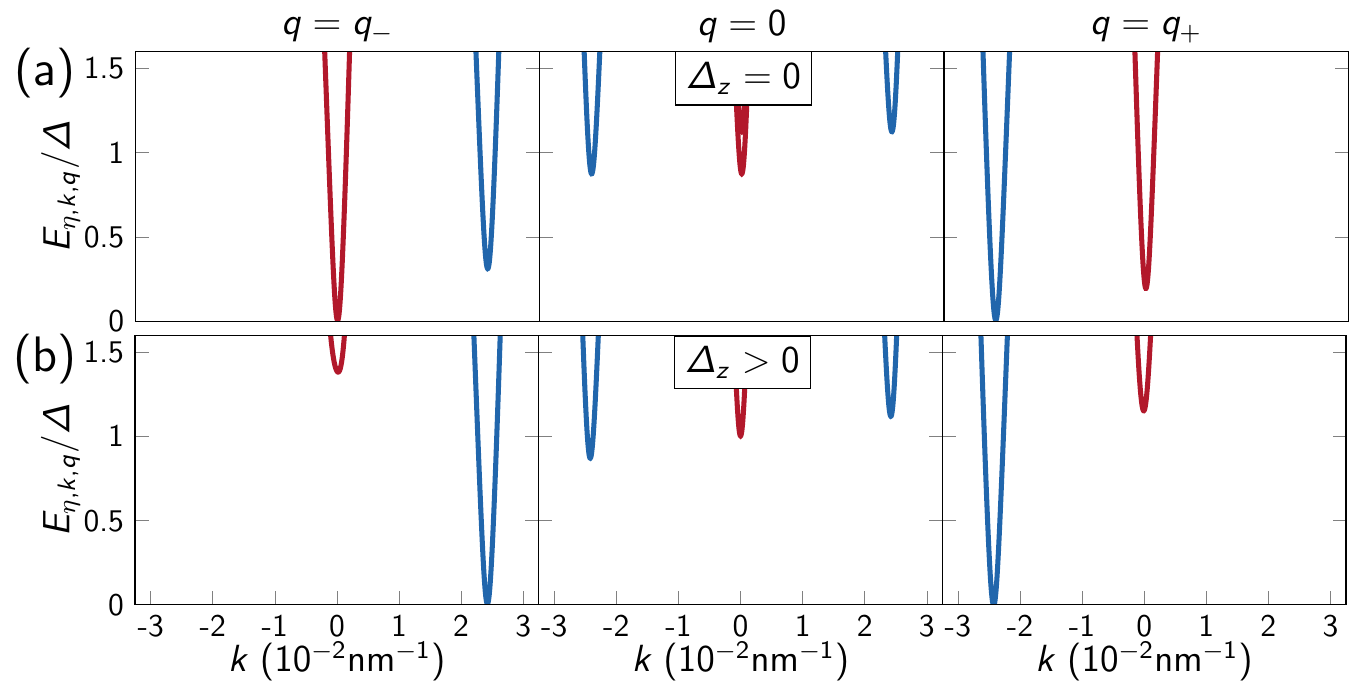}
	\caption{{\bf Finite pairing-momentum SC energy spectra of a TI nanowire at  $\mu=\mu_c$:}  Spectra 	become gapless above a positive ($q_+$) or below a negative ($q_-$)  critical momentum (right and left panels, resp.). 
	{(a)}~A magnetic field applied parallel to the spin-polarization axis 
	($\Delta_x\neq0$) results in an MCA 
	such that $q_+\neq|q_-|$.
	 Note that the interior gap (red) closes for $q_-$, whereas the exterior gap (blue) closes for $q_+$.  ({b}) In contrast, for sufficiently large $\Delta_z$, 
	due to modification of the interior gap, 
	 the critical momenta $q_+$ and $q_-$, at which the system becomes fully gapless, both correspond to the closing of the exterior gap and the SC diode effect is substantially altered.
	Parameters for all plots: $\hbar v=300$ meV nm$\approx 5\times10^5$ $\hbar$m/s, $R=15$ nm, $\delta\varepsilon=\hbar v/R=20$ meV, $\Delta_x=0.025$ meV, $\delta\mu_{2\ell}=-0.2\delta\varepsilon$, and $\Delta=0.2$ meV \cite{Munning2021,Legg2022MCA}. 
	Plot specific parameters: (a) $\Delta_z=0$, $q_-=-0.00144$~nm$^{-1}$, and $q_+=0.00155$~nm$^{-1}$. (b) $\Delta_z=0.4$, $q_-=-0.001975$~nm$^{-1}$, and $q_+=0.00155$~nm$^{-1}$.}
\label{fig2}
\end{figure}

\begin{figure*}[t]
	\centering
  \includegraphics[width=\textwidth]{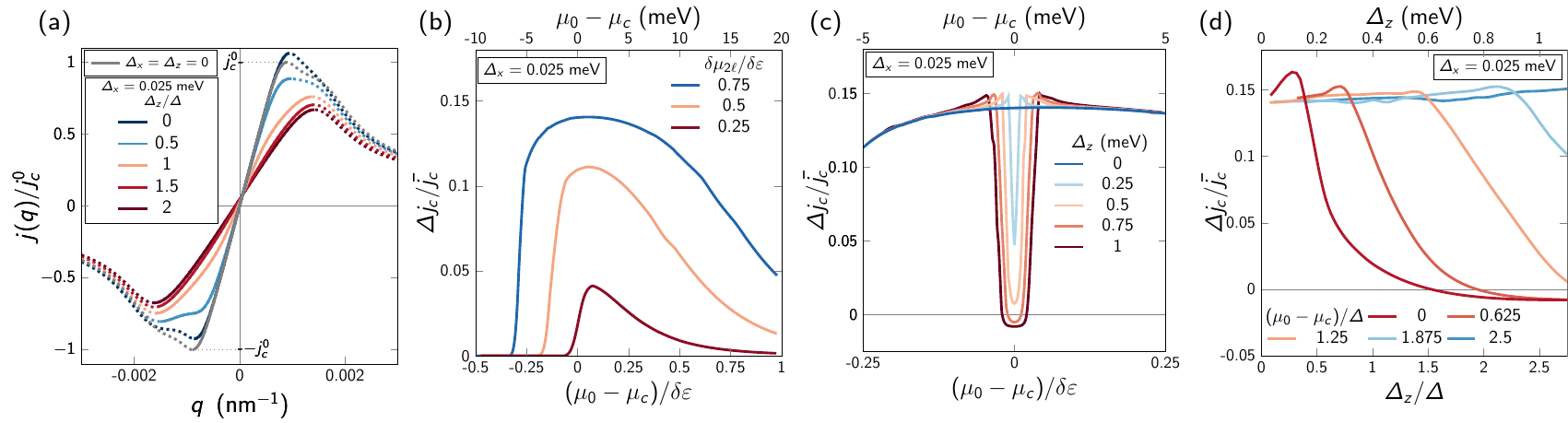}
	\caption{{\bf SC diode effect in superconducting TI nanowires:} (a) Supercurrent as a function of pair momentum, $j(q)$, at the subband crossing point $\mu_0=\mu_c$. The maximum (minimum) of $j(q)$ correspond to the extremal supercurrents $j_c^+$  ($j_c^-$). For $\Delta_x=0$ (gray), the critical current $j_c^0$ is reciprocal, i.e. $j_c^+=|j_c^-|$, however, in the presence of broken inversion symmetry at finite Zeeman energy, $\Delta_x$, the extremal supercurrent becomes non-reciprocal, i.e. $j_c^+\neq |j_c^-|$. The non-reciprocity is further modified by $\Delta_z$. (b) The SC diode quality factor, $\Delta j_c/\bar{j_c}$, as a function of chemical potential for finite $\Delta_x$ and $\Delta_z=0$. The maximal SC diode quality factor is achieved close to $\mu_0=\mu_c$. The SC diode quality factor is proportional to the non-uniformity of chemical potential, $\mu_{2\ell}$, which governs the size of the subband splitting.  (c) When, in addition $\Delta_z\neq0$, close to the subband crossing at $\mu_0\approx \mu_c$, the  quality factor is diminished. This can be used to ascertain the location in parameter space where a topological phase transition is expected to occur.  (d) The dependence of the quality factor on Zeeman energy, $\Delta_z$, for various chemical potentials close to $\mu_0=\mu_c$. A strong dependence on $\Delta_z$ indicates that $\mu$ is close to the subband crossing. Parameters: $\hbar v=300$ meV nm$\approx 5\times10^5$~$\hbar$m/s, $R=15$ nm, $\delta\varepsilon=\hbar v/R=20$ meV, $\beta^{-1}=0.05$ meV/k$_B \approx 0.5$ K, $\delta\mu_{2\ell}=-0.75\delta\varepsilon$, and $\Delta=0.4$ meV.}
	\label{fig3}
\end{figure*}

Recently it has been shown that the SC diode effect can be described phenomenologically using a finite pairing-momentum \cite{He2022,Daido2022,Noah2022} and we will follow a similar formulation here. While some aspects may not be captured by this treatment, these theories of the SC diode effect are sufficient to understand the behaviour expected in the nanowire systems considered here. An effective pairing potential, $\Delta$, in a nanowire is induced via proximity effect with a bulk superconductor. In the presence of the supercurrent, the Cooper pair has a finite momentum $q$, and the effective Hamiltonian reads
\begin{align}
H(q,\Delta)&=\frac{1}{2}\sum_k {\boldsymbol \psi}_k^\dagger \left(\begin{array}{cc}
h_{k+\frac{q}{2}} & -i \sigma_{y} \Delta \\
i \sigma_{y} \Delta & -h_{-k+\frac{q}{2}}^{*}
\end{array}\right) {\boldsymbol \psi}_k.
\end{align}
Here, we work in the basis ${\boldsymbol \psi}^\dagger_k$=($c^\dagger_{k+\frac{q}{2} \uparrow}$, $c^\dagger_{k+\frac{q}{2} \downarrow}$, $c_{-k+\frac{q}{2} \uparrow}$, $c_{-k+\frac{q}{2} \uparrow}$), where the  annihilation operator $c_{k \sigma}$ act on an electron with momentum $k$ and spin $\sigma=\uparrow$, $\downarrow$. 
Applying a Bogoliubov transformation, we arrive at 
\begin{align}
H(q,\Delta)&=\sum_{k,\eta=\pm} \left( E_{\eta,k,q} \gamma^\dagger_{\eta,k,q} \gamma_{\eta,k,q} - \frac{1}{2} E_{\eta,k,q}\right),
\end{align}
where the index $\eta$ labels two branches of the spectrum with positive energies $E_{\eta,k,q}$. If $\Delta_z=0$, spin is a good quantum  number and we get an explicit solution
\begin{equation}
E_{\pm,k,q}=\left|\sqrt{\Delta ^2+(\delta \alpha \pm \bar{\xi} )^2}\pm(\bar{\alpha} +\Delta_x)+\delta \xi\right|,
\end{equation}
where we have defined 
$\bar{\xi}=\xi_{k+\frac{q}{2}}+\xi_{-k+\frac{q}{2}}$, $\delta\xi=\xi_{k+\frac{q}{2}}-\xi_{-k+\frac{q}{2}}$, $\bar{\alpha}=\alpha_{k+\frac{q}{2}}+\alpha_{-k+\frac{q}{2}}$, and $\delta\alpha=\alpha_{k+\frac{q}{2}}-\alpha_{-k+\frac{q}{2}}$.

In order to obtain the current we start with the free energy density $\Omega(\Delta,q)$ which, up to constants, is given by \cite{Daido2022,Kinnunen2018}
\begin{equation}
\Omega(\Delta,q)=-\frac{1}{L}\sum_{k,\eta}  \frac{1}{\beta} \ln\left[\cosh(\beta E_{\eta,k,q}/2)\right],\label{freeE}
\end{equation}
where $L$ is the length of the nanowire, and $\beta=(k_{\rm B} T)^{-1}$ with $k_B$ the Boltzmann constant and $T$ temperature. Introducing the vector potential, $\vec A$, allows us to calculate the uniform supercurrent density along the nanowire such that~\cite{Noah2022}
\begin{align}
j(q)&=-\frac{\partial \Omega(\Delta,q-2e A_x)}{\partial A_x}\bigg|_{A_x=0}=2e \partial_q \Omega(\Delta, q)\label{currentden}\\
&=-\frac{e}{L}\sum_{k,\eta} \tanh(\beta E_{\eta,k,q}/2) \partial_q  E_{\eta,k,q} ,\nonumber
\end{align}
where $e<0$ is the charge of the electron and the sign of $j(q)$ indicates the direction of current flow. Since we consider here systems where the pairing potential is induced by proximity to a superconductor, we make the simplifying approximation that the induced pairing, $\Delta$, is a constant in the free energy. We find that the exact size of $\Delta$ has a limited impact on our results (see SM~\cite{SI}) and so to maintain clarity throughout we do not model the reduction in $\Delta$ due to a magnetic field.

At zero temperature, for $\Delta_z=0$, the behaviour of $j(q)$ can be read off directly from Eq.~\eqref{currentden}. For all pairing momenta where the energy spectrum remains gapped, we have $ \tanh(\beta E_{\eta,k,q}/2) =1$ and the current depends linearly on $q$, such that $j(q)=j_0 q$, with $j_0$ a constant. However, the spectrum becomes fully gapless for any $q$ above the positive ($q_+$) or below the negative ($q_-$) critical pairing momentum [see Fig.~\ref{fig3}(a)] meaning that $\tanh(\beta E_{\eta,k,q}/2)=0$ at certain values of momentum $k$, as a result $j(q)$ deviates from the linear behaviour which indicates the breakdown of superconducting phase at this momentum \cite{He2022,Daido2022,Noah2022}. 

Turning now to the SC diode effect, if $\Delta_x=0$, due to the time-reversal symmetry,  there is no MCA of the energy spectrum and the critical momenta satisfy $q_{+}=|q_{-}|$, as such the critical currents are also reciprocal, $j_c^+= |j_c^-|$. If $|\Delta_x|>0$,  it is possible to get $q_+\neq|q_{-}|$ such that the critical current becomes non-reciprocal, $j_c^+\neq |j_c^-|$, with an SC diode quality factor at zero-temperature
 \begin{equation}
\frac{\Delta j_c}{\bar{j_c}}\equiv\frac{j_c^+- |j_c^-|}{(j_c^++ |j_c^-|)/2}=\frac{q_+- |q_-|}{(q_++ |q_-|)/2}.\label{qualityfac}
\end{equation}

Furthermore, in the vicinity of the subband crossing point, an additional component of magnetic field parallel to the nanowire, $|\Delta_z|>0$, modifies the interior gap and, as a result, the critical momenta $q_{+}$ and $q_{-}$ are dependent on $\Delta_z$. More precisely we use $q_{+}$ and $q_{-}$ to refer to the momenta at which superconductivity in the nanowire becomes fully gapless for all $q$ greater (smaller) than $q_+$ ($q_-$), indicating the complete breakdown of any superconductivity in the nanowire. Due to the modification of the interior gap by $\Delta_z$ within the vicinity of the subband crossing, for sufficiently large $\Delta_z$, both $q_{+}$ and $q_{-}$ always correspond to a closing of the exterior gap [see Fig.~\ref{fig2}(b)], which can be expected to substantially modify the SC diode effect in this region of phase space. As such, the dependence of the SC diode effect on $\Delta_z$ can serve as a signature of the modification of the interior gap by a component of magnetic field parallel to the nanowire and, therefore, can be used to determine that a device is in the region of parameter space where topological superconductivity is expected.

{\it SC diode effect in topological insulator nanowires.} We consider TI nanowires as a first example for the SC diode effect behavior described above. In a TI nanowire of radius $R$ with a non-uniform potential of the form $\mu(\phi)=\mu_0+2\sum_{n> 0} \cos(n\phi)\delta \mu_n$ induced  through the cross-section -- for example, by the application of a gate or by contact with a superconductor \cite{Legg2021,Legg2022metal} -- the normal state Hamiltonian can be approximated by a form equivalent to Eq.~\eqref{eq:genham} such that~\cite{Legg2021}
\begin{equation}
\xi_k=\hbar v\sqrt{k^2+\frac{\ell^2}{R^2}}-\mu_0\;\; {\rm and} \;\; \alpha_k= \frac{\delta\mu_{2\ell}\:k}{\sqrt{k^2+\ell^2/R^2}},\label{TIsubbands}
\end{equation}
where $\mu_0$ is the average chemical potential and $l$ is a half-integer. In what follows, we consider the lowest subband pair with crossing point at chemical potential $\mu_c=\delta\varepsilon/2=\hbar v/2R$.

In order to investigate the SC diode behavior of TI nanowires, we numerically calculate the free energy at finite temperature, as described in Eq.~\eqref{freeE}, and, using Eq.~\eqref{currentden}, obtain the maxima ($j^+_c$) and minima ($j^+_c$) of $j(q)$, which corresponds to the extremal supercurrent after which the superconductivity in the nanowire becomes fully gapless \cite{He2022,Daido2022,Noah2022}. In the topological transition regime an accidental gap closing can also occur at the supercurrent induced transition from trivial to topological superconductivity \cite{Romito2012,Liu2013,Sticlet2013,Mahyaeh2018,Dmytruk2019}, however, we focus on the maximal supercurrent 
the proximitized nanowire can maintain. As expected from the preceding discussion and Fig.~\ref{fig2}(a), we find that the MCA of the energy spectrum results in an SC diode effect when superconductivity is induced via the proximity effect and a magnetic field is applied parallel to the spin-polarization axis ($\Delta_x \neq 0$) [see Fig.~\ref{fig3}(a-b)]. For realistic parameters of experimental devices~\cite{Munning2021,Legg2022MCA}, we find that the SC diode quality factor is proportional to the size of the subband splitting induced by a non-uniform chemical potential, $\delta \mu_{2\ell}$, and can easily reach $\sim$15\% difference in critical currents.  The observation of an SC diode effect of this magnitude would therefore be a strong indicator
of the large subband splitting possible in a proximitized TI nanowire device.

Next, as shown in Fig.~\ref{fig3}(c-d), we consider the impact of an additional magnetic field component parallel to the nanowire axis $\Delta_z\neq 0$. In TI nanowires, orbital effects lead to a very large $g_z$ for the effective $g$ factor along the nanowire axis, which can be useful for achieving topological superconductivity, when the chemical potential lies close to the subband-crossing at $\mu_0=\mu_c$ \cite{Legg2021}. The additional Zeeman component $\Delta_z$ drastically alters the SC diode effect in the vicinity of the subband crossing point close to $\mu_0=\mu_c$. However, away from $\mu_0=\mu_c$ the SC diode quality factor remains largely unchanged as a result of finite $\Delta_z$. In particular, we find that a large reduction in the SC diode quality factor $\Delta j/\bar{j}_c$ indicates the modification of the interior gap. The dependence 
of the SC diode 
quality factor can therefore be used to ascertain that the nanowire is in the region of phase space, where topological superconductivity is expected.

\begin{figure}[t]
	\centering
  \includegraphics[width=1\columnwidth]{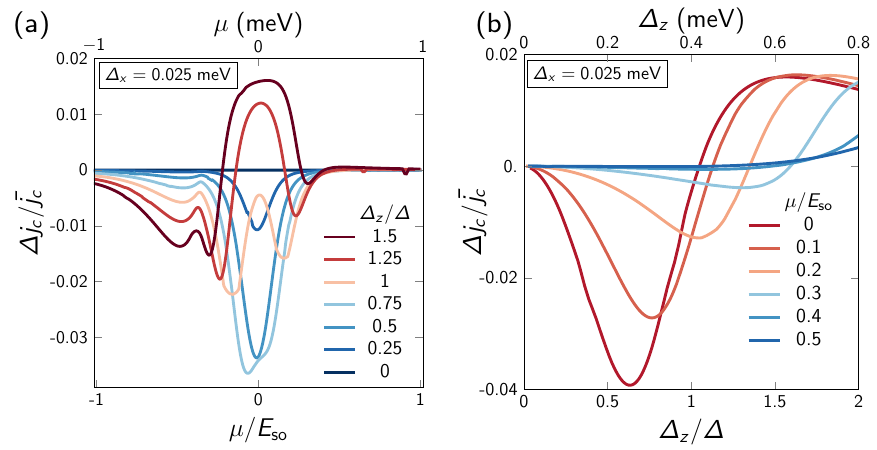}
	\caption{{\bf SC diode effect in superconducting Rashba nanowires:}  (a) The SC diode effect quality factor, $\Delta j_c/\bar{j_c}$, as a function of chemical potential $\mu$ for various $\Delta_z$: a large change in  quality factor due to $\Delta_z$ can be used as an indicator the system is close to the subband crossing  and therefore in the parameter regime where topological superconductivity is expected.  
	(b) Quality factor, $\Delta j_c/\bar{j_c}$, as a function of magnetic field for various 
	$\mu$. 
	Close to the subband crossing  at $\mu=0$, a strong dependence on $\Delta_z$ develops. In contrast, for $\mu$
	detuned from this crossing there is only a very weak dependence on $\Delta_z$. Parameters are broadly those expected in the normal state~\cite{Lutchyn2018}: $m^*_0=0.015 m_e$, $\beta^{-1}=0.05$ meV/k$_{\rm B} \approx 0.5$ K, $\alpha$=1 eV\AA, $E_{\rm so}= \frac{\alpha^2 m^*_0}{2\hbar^2}\approx 1$ meV, $\mathcal{C}$=0, $\Delta_x=0.025$ meV, and $\Delta=0.4$ meV.}
	\label{fig4}
\end{figure}

{\it SC diode effect in Rashba nanowires:} Next, we consider the SC diode effect in Rashba nanowires. In this case, the elements of the Hamiltonian in Eq.~\eqref{eq:genham} take the form 
\begin{equation}
\xi_k=\frac{\hbar^2 k^2}{2m^*(k)}-\mu\quad {\rm and} \quad \alpha_k=\alpha k,\label{rashba}
\end{equation}
where $m^*(k)$ is the (momentum dependent) effective mass and $\alpha$ is the Rashba SOI coefficient. If $\Delta_z=0$ and the effective mass $m^*(k)=m_0^*$ is constant, no MCA rectification or SC diode effect occurs because all dependence on the SOI strength, $\alpha$, as in Eq.~\eqref{rashba}, can be removed via the spin-dependent gauge transformation \cite{Braunecker2010}. However, this gauge transformation is not possible if $\Delta_z\neq0$ and/or momentum dependence of the mass is included (e.g. $\frac{\hbar^2}{2 m^*(k)}\approx \frac{\hbar^2}{2 m^*_0} +\mathcal{C} k^2$, with $\mathcal{C}$ a leading order mass correction). In these cases, MCA of the energy spectrum leads to a both MCA rectification in the diffusive normal state and an SC diode effect, which are proportional to the Rashba SOI strength (see SM~\cite{SI}). A cubic spin-orbit term, such as $\alpha_k= \beta k^3 \sigma_x$, would also lead to a SC diode effect.

The combination of Zeeman terms $\Delta_x\neq0$ and $\Delta_z\neq0$ results in a finite SC diode quality factor in the vicinity of the subband crossing point, see Fig.~\ref{fig4}. (The situation for $\mathcal{C}=0$ is considered in the SM \cite{SI}.) In particular, for small $\Delta_z$, we find that the magnitude of the SC diode quality factor $\Delta j/\bar{j}_c$ is largest at the subband crossing $\mu=0$. In the vicinity of the subband crossing, at large $\Delta_z$ and $E_{\rm so}$,
 $\Delta j/\bar{j}_c$ changes sign close to the phase transition to topological superconductivity before becoming largely independent of $\Delta_z$, see Fig.~\ref{fig4}(b). 

{\it Discussion:} We have shown that magnetochiral anisotropy of the energy spectrum in TI nanowires or Rashba nanowires results in an SC diode effect when a pairing potential is induced via proximity with a superconductor. The size of the SC diode effect can provide a measure of the magnitude of inversion symmetry breaking in the presence of superconductivity. Finally, a strong dependence of the SC diode quality factor on the magnetic field component parallel to the nanowire can be used as an indicator that the system is in the region of phase space close to where topological superconductivity and associated MBSs should occur. 

\begin{acknowledgments}
We thank Yoichi Ando and Georg Angehrn for useful conversations. This work was supported by the Georg H. Endress Foundation, the Swiss National Science Foundation, and NCCR QSIT (Grant number 51NF40-185902). This project received funding from the European Union’s Horizon 2020 research and innovation program (ERC Starting Grant, Grant No 757725).
\end{acknowledgments}
\bibliography{TI-nws}

\end{document}

\title{Supplemental Information:\\Superconducting diode effect due to magnetochiral anisotropy\\ in topological insulator and Rashba nanowires}
\author{Henry F. Legg}
\affiliation{Department of Physics, University of Basel, Klingelbergstrasse 82, CH-4056 Basel, Switzerland}
\author{Daniel Loss}
\affiliation{Department of Physics, University of Basel, Klingelbergstrasse 82, CH-4056 Basel, Switzerland}
\author{Jelena Klinovaja}
\affiliation{Department of Physics, University of Basel, Klingelbergstrasse 82, CH-4056 Basel, Switzerland}
\maketitle
\vspace{-10pt}

\begin{figure*}[th]
	\centering
  \includegraphics[width=0.55\textwidth]{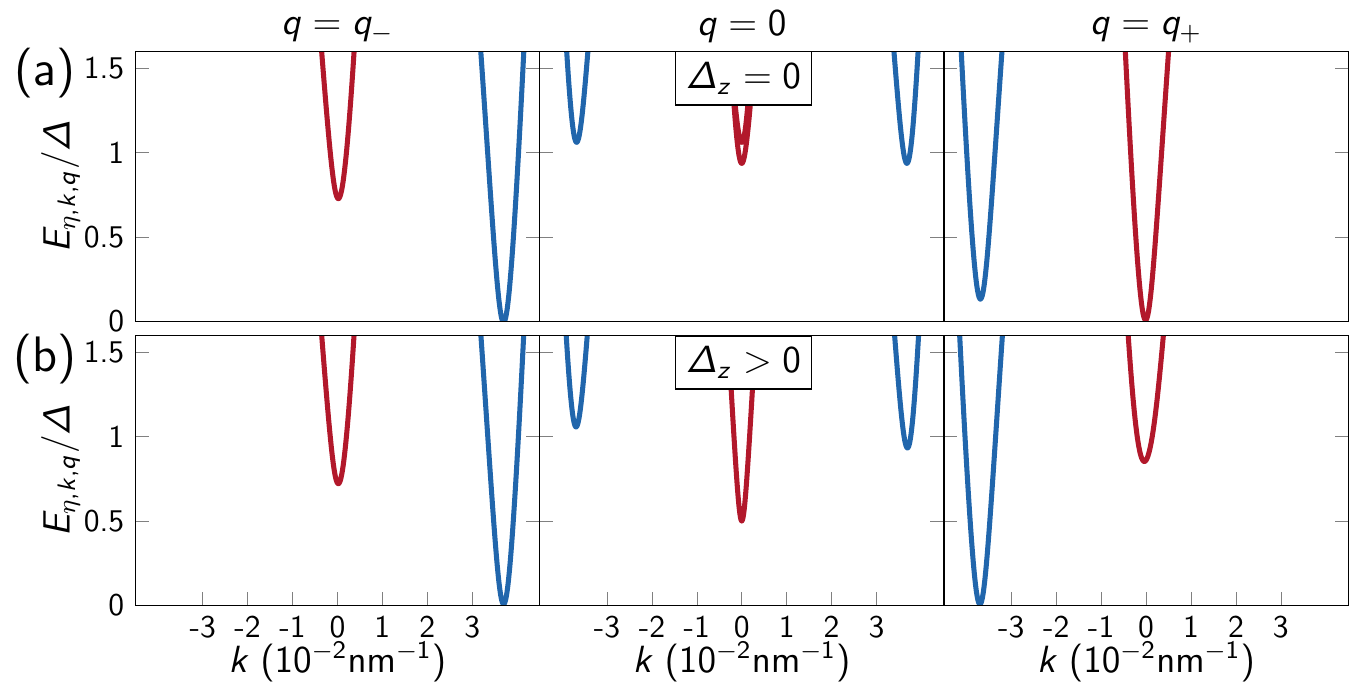}
	\caption{{\bf Finite pairing-momentum SC energy spectra of a Rashba nanowire:}  Spectra at the subband crossing point ($\mu=0$)  become fully gapless above a positive ($q_+$) or below a negative ($q_-$) finite critical momentum (right and left panels, respectively). These critical momenta correspond to the complete breakdown of the SC state. {(a)}~For the Rashba nanowire, when there is no component of magnetic field along the nanowire, $\Delta_z=0$, and the effective mass is independent of momentum, $m^*(k)=m_0^*$, then all dependence on the SOI strength, $\alpha$, can be removed via the spin-dependent gauge transformation~\cite{Braunecker2010} and as a result the critical momentum $q_+=|q_-|$ even in the presence of a finite $\Delta_x$. ({b}) In contrast, for a finite $\Delta_z$, due to the modification of the interior gap, the critical momenta $q_+\neq |q_-|$ and an SC diode effect occurs. In particular, in the topological phase the momenta $q_+$ and $q_-$ correspond to the closing of the exterior gap. Parameters for all plots: Same as Fig.~4 of main text. Plot specific parameters: (a) $\Delta_z=0$, $q_-=-0.00372$~nm$^{-1}$, and $q_+=0.00372$~nm$^{-1}$. (b) $\Delta_z=0.6$ meV, $q_-=-0.00367$~nm$^{-1}$, and $q_+=0.00417$~nm$^{-1}$.}
	\label{scplots-si}
\end{figure*}

As discussed in the main text, the critical current is related to the finite pairing momenta, $q$, where the system becomes fully gapless. Namely this occurs above a positive, $q_+$, or below a negative, $q_-$, critical momentum. If the magnitude of these momenta differ, then it can be expected that an SC diode effect will occur. In Fig.~\ref{scplots-si}, we show the critical momenta for the Rashba nanowire, equivalent to Fig.~1 of the main text for TI nanowires. In contrast to TI nanowires, however, when there is no component of magnetic field along the nanowire, $\Delta_z=0$, and the effective mass is independent of momentum, $m^*(k)=m_0^*$, even in the presence of a finite $\Delta_x$ the critical momenta still satisfy $q_+=|q_-|$. This is because all dependence on the Rashba SOI interaction strength, $\alpha$, can be removed via a gauge transformation in an open system~\cite{Braunecker2010} such that $k \rightarrow k\mp \alpha/2$ and redefinition $\mu\rightarrow \mu+\alpha^2/4$, this maps the Rashba nanowire onto a system with no inversion symmetry breaking and therefore no MCA is present. The addition of a finite $\Delta_z$ and/or the full momentum dependence of the mass, $m^*(k)$, will result in MCA of the spectrum and a SC diode effect (see below).

\section{SC diode effect in superconducting Rashba nanowires:\\Momentum dependent mass and reduced parameters}

In this section, we consider the SC diode effect in semiconductor nanowires with strong Rashba SOI and include the $k$-dependence of the full effective mass, $m^*(k)$, to leading order  such that is is approximated as $\frac{\hbar^2}{2 m^*(k)}\approx \frac{\hbar^2}{2 m^*_0} +\mathcal{C} k^2$. The components of the Hamiltonian as in Eq.~(1) of the main text then take the form
\begin{equation}
\xi_k=\frac{\hbar^2 k^2}{2m^*_0}+\mathcal{C} k^4-\mu\quad {\rm and} \quad \alpha_k=\alpha k \label{rashba}.
\end{equation}
%where $\alpha$ is the Rashba SOI strength.

\begin{figure*}[!th]
	\centering
  \includegraphics[width=0.55\textwidth]{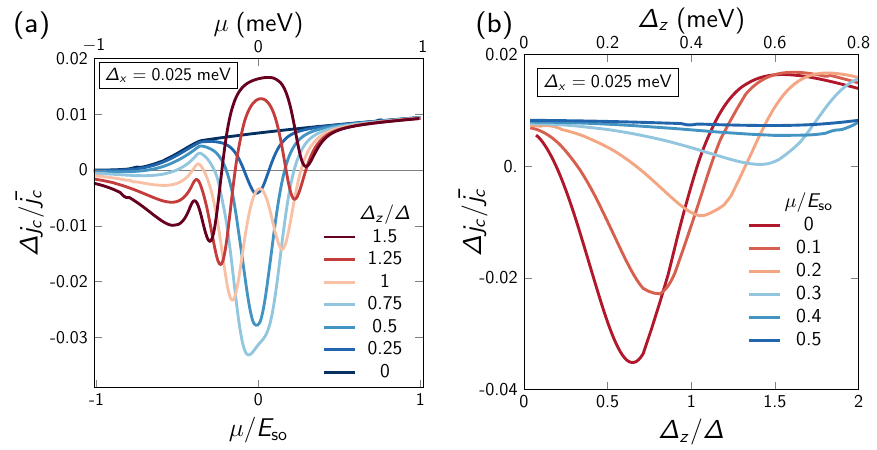}
	\caption{{\bf SC diode effect in superconducting Rashba nanowires with momentum-dependent mass:}  (a) The SC diode effect quality factor, $\Delta j_c/\bar{j_c}$, as a function of chemical potential $\mu$ (scaled by the SOI energy $E_{\rm so}$) for various $\Delta_z$: Due to $\mathcal{C}\neq0$ there is a finite SC diode effect already for $\Delta_z=0$. However, for finite $\Delta_z$, the quality factor evolves in a similar manner to that found in Fig.~4(a) of the main text where $\mathcal{C}=0$. (b) The SC diode effect quality factor, $\Delta j_c/\bar{j_c}$, as a function of magnetic field for various chemical potentials $\mu$. Apart from a finite value at $\Delta_z=0$, the behavior of the SC diode quality factor is similar to that found in Fig. 4(b) of the main text. Parameters same as Fig.~4 of main text, but with $\mathcal{C}=20000$~meV~nm$^{4}$ such that there is a $\sim 1\%$ difference in effective mass, $m^*(k)$, at $k=0$ and the SOI momentum $k=k_{\rm so}$.}
	\label{rnw-si}
\end{figure*}

\begin{figure*}[th]
	\centering
  \includegraphics[width=0.55\textwidth]{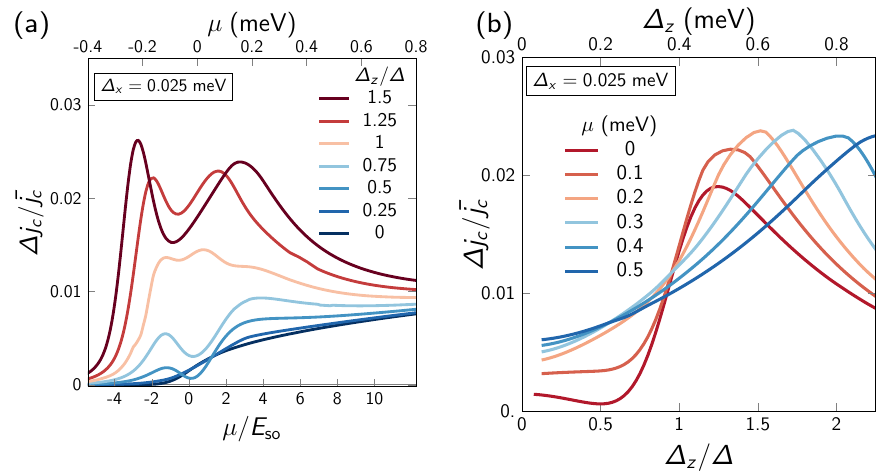}
	\caption{{\bf SC diode effect in superconducting Rashba nanowires with reduced SOI energy and enhanced effective mass:}  (a) The SC diode effect quality factor, $\Delta j_c/\bar{j_c}$, as a function of chemical potential $\mu$ for various $\Delta_z$. The reduced SOI strength $\alpha$ means that there is a reduced magnitude of the SC diode quality factor in the vicinity of the subband crossing at $\mu=0$ and there is no change of sign of the quality factor as a function of $\Delta_z$. (b) The  quality factor, $\Delta j_c/\bar{j_c}$, as a function of magnetic field for various chemical potentials $\mu$. The dependence of the quality factor on $\Delta_z$ is weaker compared to systems with large SOI energy and the sign of the quality factor does not change, even close to the subband crossing point at $\mu=0$. Parameters same as Fig.~\ref{rnw-si} apart from $\alpha$=0.2 eV\AA\;  and $m^*_0=0.03 m_e$.}
	\label{rnw-si-2}
\end{figure*}

When $\mathcal{C}\neq0$ is included, we find a SC diode effect already for $\Delta_z=0$ such that in the experimentally relevant regime the SC diode quality factor follows  (see also Sec.~\ref{dependencies} below),
\begin{equation}
\Delta j_c/\bar{j_c}\propto B_x \alpha \mathcal{C}.
\end{equation}
%Experimentally 
Using parameters as in the main text, but with a realistic finite $\mathcal{C}$ gives a SC diode quality factor smaller than in TI nanowires but still $\Delta j_c/\bar{j_c}\sim 1\%$ for $\Delta_x\neq0$, see Fig.~\ref{rnw-si}. In Fig.~\ref{rnw-si-2}, we consider a reduced value of the RashbaSOI strength $\alpha$ and an increased mass $m_0^*$ in order to consider the impact of metallization on the nanowire by the superconductor. We find that the overall dependence on $\Delta_z$ is much weaker than in the case of strong SOI, as in Fig.~4 of the main text, and there is no change of sign of the diode quality factor for these parameters.

\section{Parameter dependencies of SC diode quality factor}\label{dependencies}
\begin{figure*}[t]
	\centering
  \includegraphics[width=0.55\textwidth]{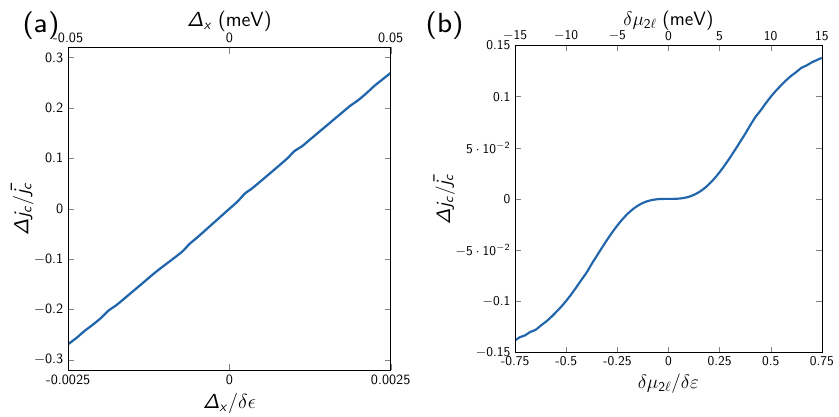}
	\caption{{\bf Dependence of SC diode effect in TI nanowires on magnetic field and non-uniform potential:}  The SC diode quality factor as a function of (a) Zeeman energy, $\Delta_x$, and (b)  the non-uniformity of chemical potential through the TI nanowire cross-section, as governed by $\delta\mu_{2\ell}$, see main text. At weak magnetic field strengths, the quality factor is linear in $\Delta_x$. In the experimentally relevant parameter range where non-uniformity of chemical potential between top and bottom surface are a few meV \cite{Legg2021,Legg2022MCA}, the   SC diode quality factor also grows approximately linearly with the non-uniformity coefficient $\delta\mu_{2\ell}$. Parameters are the same as in Fig.~3(c-d) of main text, with $\mu_0=0.75\delta \varepsilon$.}
	\label{TI-si}
\end{figure*}

\begin{figure*}[t]
	\centering
  \includegraphics[width=0.75\textwidth]{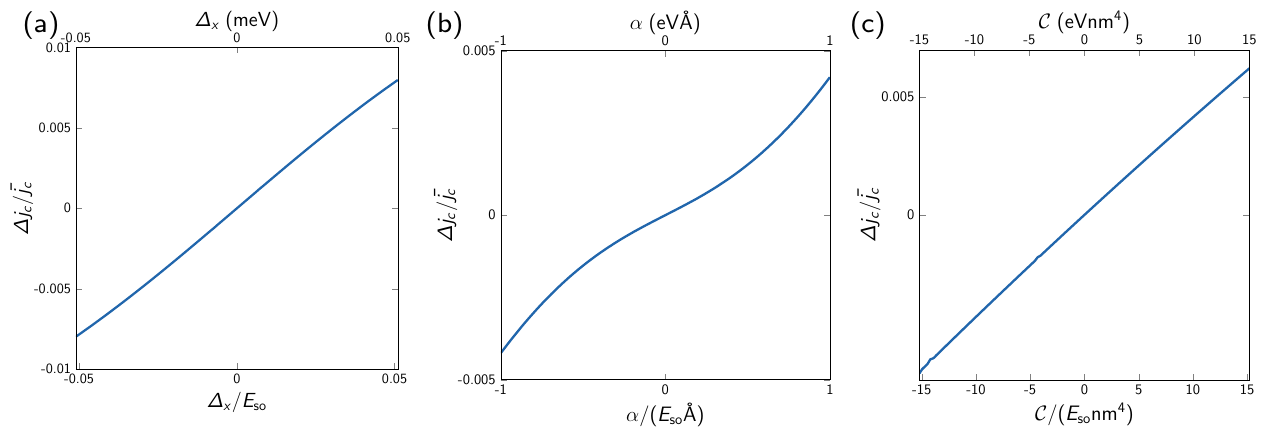}
	\caption{{\bf Dependence of SC diode effect in Rashba nanowires on various parameters:} The SC diode quality factor, $\Delta j/\bar{j}_c$, as a function of (a) Zeeman energy $\Delta_x$, (b) the SOI strength $\alpha$, and (c) the mass correction term $\mathcal{C}$. For weak magnetic field strengths, it depends linearly on $\Delta_x$.  For SOI strengths $\alpha \lesssim 0.5$ eV$\rm \AA$,  the quality factor also grows approximately linearly.  In addition, the size of the SC diode effect grows linearly with increasing $\mathcal{C}$. As expected from the dependence of the MCA rectification, the SC diode vanishes in the case $\mathcal{C}=0$ since the subband curvature $\mathcal{V}^\eta(k)=1/m$ is a constant in this case. Parameters are the same as in main text Fig.~4 with $\mu=0.5 E_{\rm so}$, apart from $\mathcal{C}=10000$ eV nm$^{4}$ in (a) and (b) .}
	\label{rnw-si-dep}
\end{figure*}

\begin{figure*}[thb]
	\centering
  \includegraphics[width=1\textwidth]{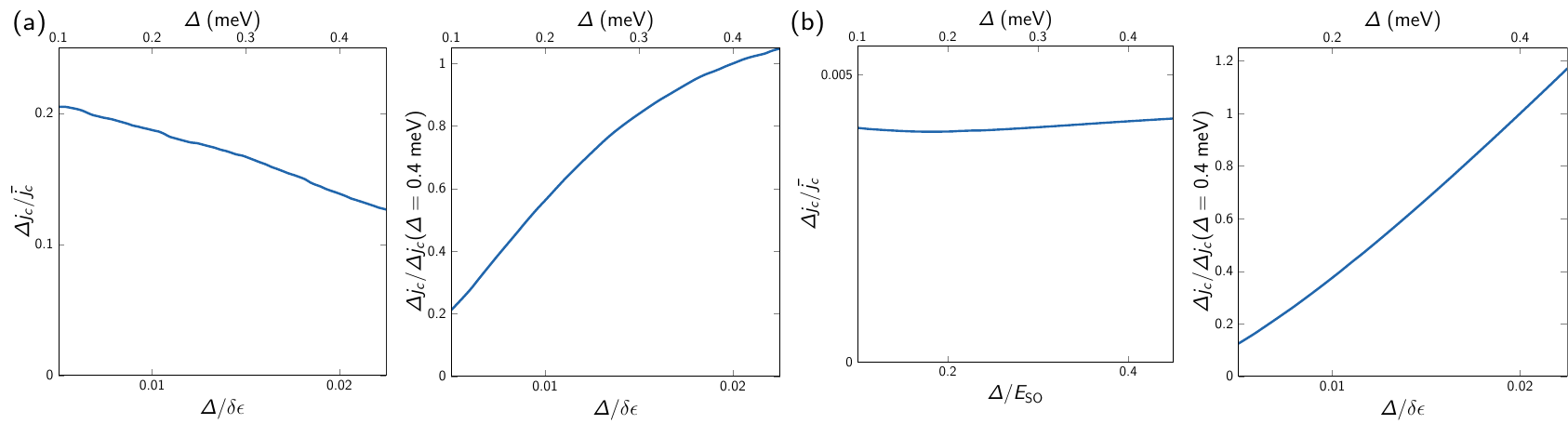}
	\caption{{\bf Dependence of SC diode quality factor and non-reciprocity of critical current on magnitude of pairing potential $\Delta$:} (a) Dependence of SC diode quality factor and magnitude of critical current non-reciprocity in TI nanowires. Left: SC diode quality factor, $\Delta j/\bar{j}_c$,  as a function of pairing potential $\Delta$. Only a very weak dependence on $\Delta$ is visible. Right: Magnitude of critical current non-reciprocity normalized by value $\Delta{j}_c$ at $\Delta=0.5$. The absolute size of the non-reciprocity $\Delta j$ does strongly depend on the induced pairing potential, $\Delta$. (b) Same dependencies  of SC diode quality factor and non-reciprocity of critical current (a), but for a Rashba nanowire. As in (a) the magnitude of the SC diode quality factor is only weakly dependent on the induced pairing $\Delta$, but the absolute size of the non-reciprocity $\Delta j$ does strongly depend on $\Delta$. Parameters: Same as in Fig.~\ref{TI-si} and Fig.~\ref{rnw-si}}.
	\label{del-si}
\end{figure*}

In this section, we consider the various parameter dependencies of the SC diode effect in TI and Rashba nanowires, as discussed in the main text. In Fig.~\ref{TI-si} we consider the case of the TI nanowire. First, in Fig.~\ref{TI-si}(a), we show that the quality factor, $\Delta j/\bar{j}_c$, depends linearly on Zeeman energy, $\Delta_x$. In Fig.~\ref{TI-si}(b), we show that the quality factor depends approximately linearly -- within an experimentally realistic range~\cite{Legg2022MCA} -- on the subband-splitting coefficient $\delta\mu_{2\ell}$. In Fig.~\ref{rnw-si}, we consider the case of the Rashba nanowire. We show that the SC diode quality factor in this case also depends linearly on Zeeman energy, $\Delta_x$, as shown in Fig.~\ref{rnw-si-dep}(a). In Fig.~\ref{rnw-si-dep}(b), we show that the SC diode quality factor depends approximately linearly on the Rashba spin-orbit strength $\alpha$. In Fig.~\ref{rnw-si-dep}(c), we show that the SC diode quality factor depends linearly on the mass correction term $\mathcal{C}$. Finally, in Fig.~\ref{del-si}, we show that  the SC diode quality factor, $\Delta j/\bar{j}_c$, for a given pairing potential $\Delta$ is only weakly dependent on the size of the pairing potential. However, we also show that the value of the non-reciprocity of the critical current, $\Delta j=j_c^+-|j_c^-|$, does depend strongly on the magnitude of the pairing potential, $\Delta$.

\newpage

\bibliography{TI-nws}